\documentclass[11pt,twoside]{article}

\usepackage{asp2006}
\usepackage{epsf}
\usepackage{psfig}
\usepackage{graphicx}
\usepackage{lscape}
\usepackage{amsbsy}
\usepackage{amssymb}

\begin{document}

\title{SOLIS Vector Spectromagnetograph: status and science}
\author{C. J. Henney$^1$, C. U. Keller$^{1,2}$, J. W. Harvey$^1$, M. K. Georgoulis$^3$, 
N.~L. Hadder$^1$, A. A. Norton$^1$, N.-E. Raouafi$^1$, R. M. Toussaint$^1$}

\affil{$^1$National Solar Observatory, Tucson, Arizona, USA}
\affil{$^2$Sterrekundig Instituut, Utrecht University, Utrecht, The Netherlands}
\affil{$^3$Johns Hopkins University APL, Laurel, MD, USA}

\begin{abstract}
The Vector Spectromagnetograph (VSM) instrument has been recording photospheric 
and chromospheric magnetograms daily since August 2003. Full-disk photospheric vector 
magnetograms are observed at least weekly and, since November 2006, area-scans of 
active regions daily. Quick-look vector magnetic images, plus X3D and FITS 
formated files, are now publicly available daily. In the near future, 
Milne-Eddington inversion parameter data will also be available and a typical 
observing day will include three full-disk photospheric vector magnetograms. 
Besides full-disk observations, the VSM is capable of high temporal cadence 
area-scans of both the photosphere and chromosphere. Carrington rotation and 
daily synoptic maps are also available from the photospheric magnetograms and 
coronal hole estimate images.
\end{abstract}

\section{Introduction}
The VSM currently operates at Kitt Peak, Arizona and is part of the Synoptic Optical 
Long-term Investigations of the Sun (SOLIS) project \citep{Keller03}. The VSM 
provides a unique record of solar full-disk vector magnetograms along with 
high-sensitivity photospheric and chromospheric longitudinal magnetograms \citep{Henney06}. 
The VSM began recording full-disk vector magnetograms in August 2003 at a temporary 
site in Tucson, Arizona. In April 2004, the VSM was relocated to Kitt Peak and 
resumed operations in May 2004. The VSM records the full Stokes profiles of the 
Fe I 630.15 and 630.25 nm lines with a current spatial and spectral sampling of 
1.125 arcsec and 2.71 pm respectively. The 50-cm aperture VSM utilizes a 
Ritchey-Chr\' etien optical design, where full-disk images are constructed from 
2048 individual steps in declination of the projected solar image on the entrance 
slit (hereafter referred to as scan-lines). Observations with fewer scan-lines 
are referred to as area-scans and are typically recorded for active 
region areas. For example, 88 area-scans were recorded in the month of October 
2003 during an exceptionally active flare period. 

The VSM is available for user-requested programs and currently supports 
ongoing solar missions (e.g. MDI, STEREO, Hinode/SOT) and will support 
forthcoming missions (e.g. SDO/HMI). For example, VSM observations provide temporal and 
spatial context for the SOT observations. Though SOT/SP has exceptional spatial resolution, 
the VSM observes any solar region eight times faster than SOT/SP and ten times larger 
in one spatial direction. The VSM also provides full-disk coverage not available 
with SOT. In addition, the future SDO/HMI instrument will have high temporal cadence 
and spatial resolution, but will have very limited spectral resolution (with six filter 
images) that the VSM will complement with fully resolved Stokes profiles.
\begin{figure*}[!ht]
\plotone{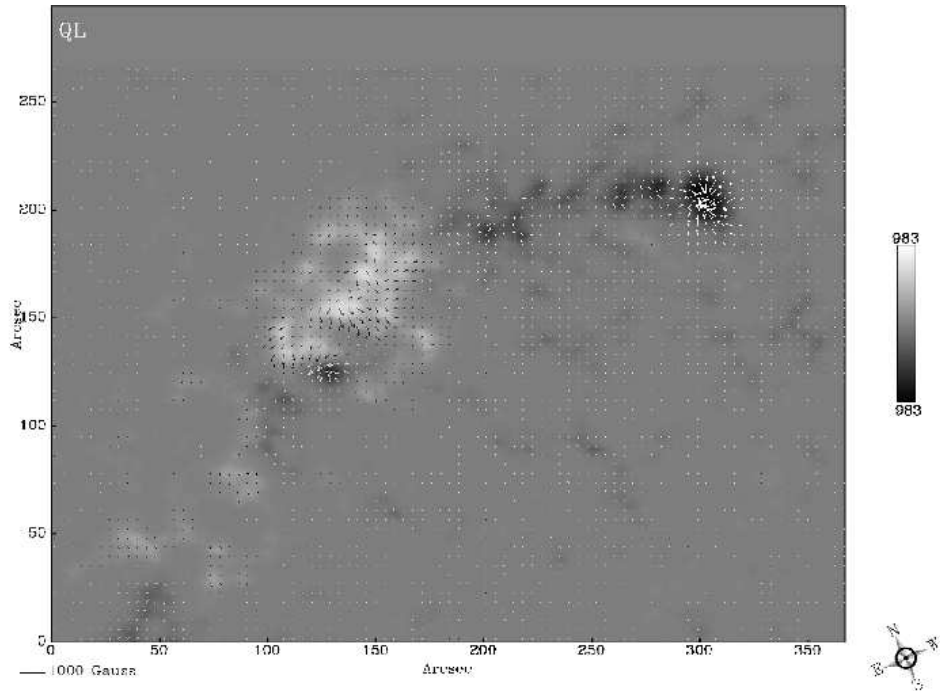}
\caption{An example VSM vector quick-look image of NOAA active region 10921, 
observed on November 3, 2006. The gray-scale illustrates the 
line-of-sight (LOS) magnetic flux and the arrows indicate the transverse field 
strength (the length scale is shown in the lower left-hand corner) and direction.}
\label{QL-AR}
\end{figure*}

\subsection{VSM Upgrades} 
Starting in 2004, the VSM suffered from a slow degradation of the polarization 
modulator needed to measure full Stokes profiles. This complicated VSM vector 
calibrations and slowed efforts to release the vector 
magnetic data. In February of 2006, the vector modulator was replaced along with
two modulators for the line-of-sight component of the magnetic field in the solar
photosphere (630.2 nm) and the chromosphere (854.2 nm). During this maintenance
period, a broken part was also replaced to reduce grating flexure. Though the new 
modulators are performing well, e.g. the strength
of the 854.2 nm signal has significantly increased, the vector modulator has a
polarization fringe pattern that is temperature dependent. The fringe pattern has 
recently been well parameterized to minimize the number of terms required for a good
fit. The fringe fitting algorithm is expected to be incorporated in the SOLIS 
processing pipeline in early 2008.
Since 2003, the VSM has incorporated two interim CMOS hybrid cameras made by 
Rockwell Scientific (now Teledyne Scientific \& Imaging, LLC). 
During the first year of operation, several unexpected 
camera signals were revealed. Two of the pronounced artifacts, a variable dark 
level and signal cross-talk at the 0.5\% level, required significant modification 
to the data reduction pipeline. The cameras were also found to have a 25\%
residual signal from previous frames. These interim cameras are scheduled to 
be replaced early in 2008 with cameras produced by Sarnoff that better match 
the spatial scale, sensitivity, low readout noise and desired frame rate for the VSM.

\section{VSM Data Products}
The raw VSM spectral data is processed using a pool of distributed data reduction 
processes that are managed by the Data Handling System on 10 dual-CPU Linux 
computers \citep[see][]{Wamp02,Jones02}. The VSM vector data processing pipeline 
is designed to provide products in two stages: first, as a quick-look (QL) product 
within 10 minutes of data acquisition, 
and then as a Milne-Eddington (ME) inversion product within 3 to 24 hours of each 
observation. The QL parameter data, following \citet{Auer77}, include 
estimates of the magnetic field strength, inclination, and azimuth \citep[see][]{Henney06}.
The ME vector parameters are determined with the inversion technique 
of \citet{Sku87}. Currently, the VSM QL data is made publicly available using 
azimuth disambiguation code that automatically selects the active regions following 
\citet{Georg08}. Once the polarization fringe fitting algorithm, discussed in 
Section 1.1, is part of the DHS, the ME vector parameter products will be publically 
available.

\begin{figure*}[!ht]
 \plotone{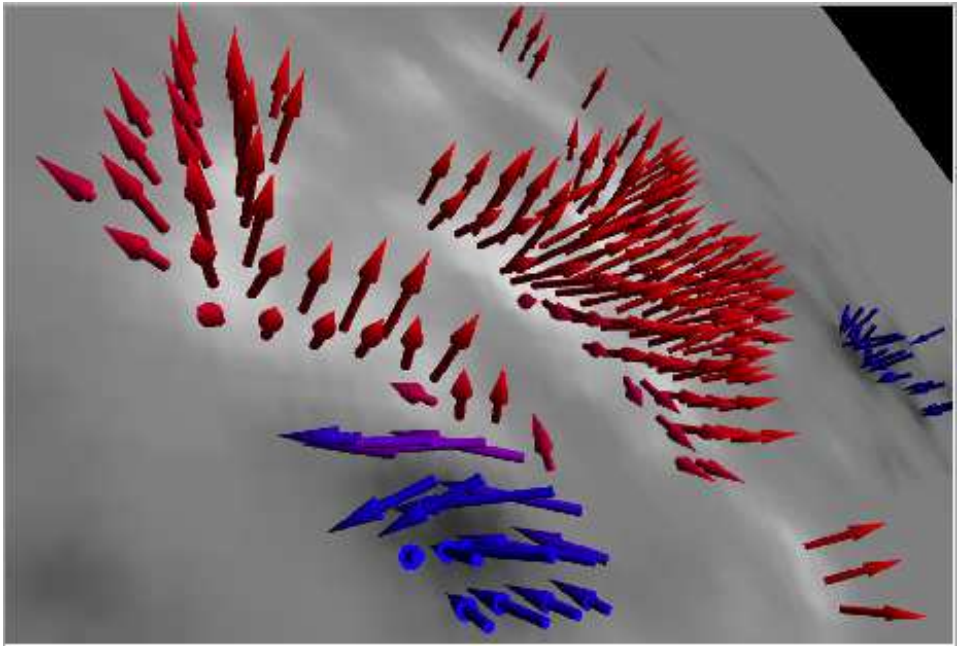}
\caption{An example snapshot of VSM quick-look X3D formatted data 
(using Flux Player by Media Machines, Inc), where the arrows indicate the 
orientation of the magnetic field. The X3D formatted 
data allows the VSM vector data to be explored as a 3-D virtual model. 
This snapshot exhibits the magnetic field orientation of active 
region AR 10921 using VSM data observed November 3, 2006.}
\label{QL-X3D}
\end{figure*}

All of the VSM QL data are corrected for the 180-degree ambiguity using the 
Non-Potential Field Calculation (NPFC) method \citep{Georg05}. VSM quick-look 
vector magnetic FITS-formatted data and JPEG 
image files are available for recent observations. An example disambiguated 
active region using VSM QL data is shown in Figure~\ref{QL-AR}. The VSM quick-look 
vector data is also available in X3D format such that the data can be explored as 
a 3-D virtual model. An example 3-D snapshot of the magnetic field for an active region, 
using VSM data, is shown in Figure~\ref{QL-X3D}. 
In addition, Carrington rotation and daily synoptic maps derived from VSM photospheric
magnetograms are available. Estimated coronal hole synoptic images and maps, derived
from He I 1083.0 nm data \citep[see][]{Henney05}, are also available.

\section{Early Science Results}
\citet{Raouafi07} utilized VSM chromospheric (Ca II 854.2 nm) 
LOS-magnetograms to study the latitude distribution of the magnetic 
flux in the polar regions. For the northern polar cap observed 
during Sep-Dec 2006, a strong latitudinal dependence
was found: increasing towards 75 deg-lat, then flatting to the pole. 
In addition, using GONG and VSM magnetogram data, \citet{Harvey07} 
found that the solar photosphere has a weak ubiquitous and 
dynamic horizontal magnetic field. 
The field variance is found strongest near the limb suggesting low-lying 
field line connections between convectively driven erupting and evolving 
network and intra-network fields. This work prompts a cautionary note on 
the common assumption that the observed photospheric weak magnetic fields are
radial when extrapolating into the corona. 

\acknowledgements
The authors acknowledge many years of dedicated effort by the SOLIS team. The
SOLIS VSM data used here are produced cooperatively by NSF/NSO and NASA/LWS. 
The National Solar Observatory is operated by AURA, Inc. 
under a cooperative agreement with the NSF.


\begin{thebibliography}{}

\bibitem[Auer, Heasley \& House(1977)]{Auer77}
Auer, L. H., Heasley, J. N., \& House, L. L. 1977, Solar Physics, 55, 47

\bibitem[Georgoulis(2005)]{Georg05}
Georgoulis, M. K. 2005, \apj, 629, L69

\bibitem[Georgoulis, Raouafi, \& Henney(2008)]{Georg08}
Georgoulis, M. K., Raouafi, N.-E., \& Henney, C. J. 2008, ASP Conf. Series, 383, 107

\bibitem[Harvey et al.(2007)]{Harvey07}
Harvey, J., Branston, D., Henney, C. J., \& Keller, C. U. 2007, \apj, 659, L177

\bibitem[Henney \& Harvey(2005)]{Henney05}
Henney, C. J. \& Harvey, J. W. 2005, ASP Conf. Series, 346, 261

\bibitem[Henney, Keller, \& Harvey(2006)]{Henney06}
Henney, C. J., Keller, C. U., \& Harvey, J. W. 2006, ASP Conf. Series, 358, 92

\bibitem[Jones et al.(2002)]{Jones02}
Jones, H. P., Harvey, J. W., Henney, C. J., Hill, F., \& Keller, C. U. 2002,
Proc. IAU Colloq. 188, ESA SP-505, 15

\bibitem[Keller, Harvey, \& Giampapa(2003)]{Keller03}
Keller, C. U., Harvey, J. W., \& Giampapa, M. S. 2003, Proc. SPIE, 4853, 194

\bibitem[Raouafi, Harvey, \& Henney(2007)]{Raouafi07}
Raouafi, N.-E., Harvey, J. W., \& Henney, C. J. 2007, \apj, 669, 636

\bibitem[Skumanich \& Lites(1987)]{Sku87}
Skumanich, A. \& Lites, B. W. 1987, \apj, 322, 473

\bibitem[Wampler(2002)]{Wamp02}
Wampler, S. 2002, Proc. SPIE, 4848, 85

\end{thebibliography}
\end{document}